\documentclass[aps,prb,superscriptaddress,twocolumn]{revtex4-1}
\usepackage{bm}
\usepackage{amsmath}
\usepackage{hyperref}
\usepackage{graphicx}
\usepackage{epstopdf}
\usepackage[english]{babel}

\begin{document}

\title{Effective Dresselhaus and Rashba spin-orbit interactions in narrow quantum wells}

\author{Zh.A. Devizorova}
\email{devizorovazhanna@gmail.com}
\affiliation{Moscow Institute of Physics and Technology, Institutskiy per. 9, Dolgoprudny, Moscow Region, 141700 Russia}
\affiliation{Kotelnikov Institute of Radio-engineering and Electronics of the Russian Academy of Sciences, 11-7 Mokhovaya St, Moscow, 125009 Russia}

\author{V.A. Volkov}
\affiliation{Kotelnikov Institute of Radio-engineering and Electronics of the Russian Academy of Sciences, 11-7 Mokhovaya St, Moscow, 125009 Russia}
\affiliation{Moscow Institute of Physics and Technology, Institutskiy per. 9, Dolgoprudny, Moscow Region, 141700 Russia}
\date{\today}

\begin{abstract}
Rashba and linear Dresselhaus interactions are believed to yield dominant contribution to the spin splitting of two-dimensional electrons in the quantum wells based on A$_3$B$_5$ compounds. We show that the interfacial spin-orbit interaction significantly renormalizes the value of the corresponding Rashba ($\alpha_{SIA}$) and Dresselhaus ($\alpha_{BIA}$) parameters. For this purpose, we solve the effective mass equation in a quantum well supplemented by the original boundary conditions on the atomically sharp interfaces and calculate the interfacial contributions to $\alpha_{SIA}$ and $\alpha_{BIA}$. Our results explain a considerable spread in the experimental data on spin-orbit parameters in GaAs/AlGaAs quantum wells. We also demonstrated that the non-equivalence of the interfaces leads to the anisotropy of the spin splitting even in quantum wells with zero average electric field.
\end{abstract}

\maketitle

\section{Introduction}
The conservation of the spin polarization is crucial for spintronic device applications. Due to the spin-orbit interaction (SOI), electrons in quantum wells (QWs) experience spin relaxation and dephasing by Dyakonov-Perel mechanism \cite {Dyakonov_FTT_1971}. In the QWs based on A$_3$B$_5$ compounds there are two types of the SOI: Dresselhaus \cite{Dresselhaus_PRB_1955} and Rashba \cite{Bychkov_JETPL_1984}. The Dresselhaus-type SOI is believed to originate from the lack of inversion symmetry in the bulk crystal and is proportional to the Dresselhaus parameter $\alpha_{BIA}$. The Rashba-type SOI is due to the structural asymmetry and is proportional to the parameter $\alpha_{SIA}$. If $\alpha_{BIA}$ and $\alpha_{SIA}$ are equal, the spin polarization of a helical spin state is conserved \cite {Koralek_Nature_2009}. To achieve this regime, one should know and control the values of both parameters. In the envelope functions approximation they are determined by the following expressions $\alpha_{BIA}^{(0)}=\gamma_c \langle \hat p_z^2 \rangle/\hbar^3$,  $\alpha_{SIA}^{(0)}=a_{so} \langle \partial_z V(z) \rangle$, where $\gamma_c$ and $a_{so}$ are bulk constants \cite{Winkler} . Thus, it is believed that the parameter $\alpha_{SIA}$ can be tuned by using the gate electrodes or by choosing the ratio between the dopant concentration on the two sides of QW, while the parameter $\alpha_{BIA}$ is determined by choosing the material and the width of QW. However, the experimental determination of the bulk constants $\gamma_c$ and $a_{SO}$ is still challenging. In spite of many experimental investigations in a wide range GaAs-based QWs, the precise value of $\gamma_c$ is still being discussed controversially in the literature. 

The parameter $\gamma_c$ was measured by Marushak with collaborators for bulk GaAs \cite{Marushchak_FTT_1983}and the value $\gamma_c=24$ eV$\times$ \AA$^3$ was obtained which is in the good agreement with $kp$-theory and has not been revised yet. But since 1990th $\gamma_c$ bas been measured not in bulk GaAs, but in QWs with the interfaces \cite{Koralek_Nature_2009, Dresselhaus_PRL_1992, Richards_SSE_1996, Miller_PRL_2003, Krich_PRL_2007, Leyland_PRB_2007, Studer_PRB_2010, Eldridge_PRB_2011, Walser_PRB_2012, Alekseev_JETPL_2013, Ganichev}. There is a considerable spread in the data obtained (see Fig.1), moreover, they are inconsistent with the theoretical results. As it was pointed out in Ref. [\onlinecite{Devizorova_PhETF_2014}], the possible reason of the spread is an incomplete account of the interfacial spin-orbit interaction (ISOI). Thus, not bulk values, but some effective quantities containing the information about the microscopic structure of the interfaces are obtained in the experiments. The theory of ISOI in the wide unilaterally doped GaAs quantum well, where the electrons are pushed toward the (001) GaAs/AlGaAs heterointerface by the built-in electric field was developed in Refs.[\onlinecite{Devizorova_PhETF_2013, Devizorova_PhETF_2014}]. The interfacial contributions to the $\alpha_{BIA}$ and $\alpha_{SIA}$ were shown to be of the same order as bulk ones. However, in a more general situation the electrons interact with atomically sharp interfacial potentials of two heterointerfaces, and ISOI at both of them contribute to  $\alpha_{BIA}$ and $\alpha_{SIA}$. To take it into account, we develop the theory of the ISOI in the QWs with an arbitrary thickness and potential profile in the present paper.

\begin{figure} 
	\includegraphics[width=6cm]{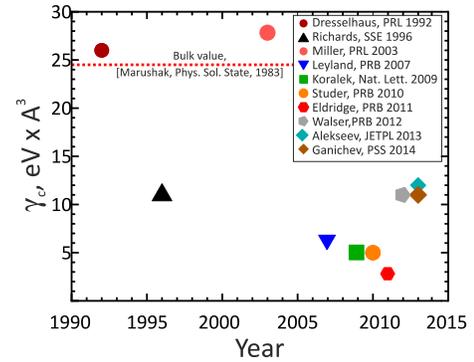}
	\caption{\label{Fig:ESspectra} The values of the bulk spin-orbit constant $\gamma_c$ extracted from the experimental data obtained by different groups in GaAs/AlGaAs quantum wells. There is a considerable spread in the data and inconsistency with the result of bulk measurements.}
\end{figure}

In the QWs grown in $z || [001]$ direction, the spin splitting of the 2D electron spectrum has the general form

\begin{equation}
\label{NQW_SS}
E_{SS}=2 p\sqrt{\alpha_{BIA}^2+\alpha_{SIA}^2+2\alpha_{BIA}\alpha_{SIA}\sin 2\phi},
\end{equation}
where $p_x=p\cos \phi$, $p_y=p\sin \phi$ are the components of 2D momentum.

To derive the interfacial contributions to $\alpha_{BIA}$ and $\alpha_{SIA}$ we begin with 3D problem in which an effective wave function $\phi$ of the conduction electron obeys the effective mass equation inside the QW of the thickness $d$. The corresponding Hamiltonian $\hat H$ contains the terms $\hat H_{BIA}$ and $\hat H_{SIA}$ describing the spin splitting of the spectrum arising due to the lack of inversion symmetry in the bulk crystal and asymmetry of the structure, respectively:
\begin{equation}
\label{H}
\hat H=\frac{{\hat p}^2}{2m^*}+V(z)+\hat H_{BIA}+\hat H_{SIA},
\end{equation}
\begin{equation}
\label{H_BIA}
\hat H_{BIA}=\frac{\gamma_c}{\hbar^3 } \biggl[\sigma_x p_x (p_y^2-\hat p_z^2)+\sigma_y p_y (\hat p_z^2-p_x^2)+\sigma_z \hat p_z (p_x^2-p_y^2)\biggr],
\end{equation}
\begin{equation}
\label{H_SIA}
\hat H_{SIA}=a_{so}(\sigma_x p_y - \sigma_y p_x)\partial_z V(z),
\end{equation}
where $\sigma_x$, $\sigma_y$ and $\sigma_z$ are the Pauli matrices.

Aiming to take into account the microscopic structure of the interfaces, we introduce appropriate boundary conditions (BCs) for the effective wave function. The phenomenological BC for a single (001) GaAs/AlGaAs heterointeface with C$_{2v}$ symmetry taking into account the spin-orbit interaction with atomically sharp interfacial potential was derived in Refs.[\onlinecite{Devizorova_PhETF_2013, Devizorova_PhETF_2014}] from the general physical requirements. Since the interfaces are, in general, non-equivalent we describe them by such BCs with different phenomenological parameters

\begin{equation}
\label{BC}
\hat \Gamma_1 \phi(z)|_{z=d/2}=0,  \qquad   \hat \Gamma_2 \phi(z)|_{z=-d/2}=0,
\end{equation}

\begin{multline}
\label{Gamma}
\hat \Gamma_{1(2)}=\Biggl. \Biggl[\hat 1 - i\frac{R_{1(2)}}{\hbar}{\bf \hat p} {\bf n} -i\frac{2m^* \gamma_c R_{1(2)}}{\hbar^4}(\sigma_y p_y -\sigma_x p_x) {\bf \hat p} {\bf n} +\\+ \frac{(\chi+\chi^{int}_{1(2)}) R_{1(2)}}{\hbar} \bm {\sigma}({\bf \hat p}\times {\bf n})- \frac{2m^*\gamma_{c_{1(2)}}^{int}}{\hbar^3}(\sigma_y p_y - \sigma_x p_x)\Biggr].
\end{multline}

Here ${\bf n}$ is the unit vector directed along the external normal to the corresponding interface; $R_1$ ($R_2$) is a real quantity describing the spectrum of the Tamm's states near the right (left) boundary if they exist (for this sake the condition $R>0$ must be fulfilled); $\chi$ is the bulk spin-orbit parameter ($\chi=0.082$ for GaAs); $\gamma_{c_{1}}^{int}$, $\chi_1^{int}$ and $\gamma_{c_{2}}^{int}$, $\chi_2^{int}$ characterizes the spin-orbit interaction at the right and left heterointerfaces, respectively. 

In the lowest order over the scalar contributions of the interfaces and the ISOI parameters, the operators $\hat \Gamma_{1(2)}$ in the BCs (\ref{BC}) can be transformed to the unitary form 
$\hat {\tilde \Gamma}_{1(2)}=\exp\left(i\hat g_{1(2)} \hat p_z / \hbar \right)$ with $\hat g_{1(2)}$ satisfying
\begin{multline}
\hat g_{1(2)}=-R_{1(2)}n_z-\frac{2m^* \gamma_c R_{1(2)}}{\hbar^3}(\sigma_y p_y -\sigma_x p_x)n_z -\\-\frac{(\chi+\chi^{int}_{1(2)}) R^2_{1(2)}}{\hbar}(\sigma_x p_y -\sigma_y p_x)-\\-\frac{2m^*\gamma_{c_{1(2)}}^{int}R_{1(2)}}{\hbar^3}(\sigma_y p_y - \sigma_x p_x)n_z,
\end{multline}

where $n_z=1$ for the right interface and $n_z=-1$ for the left one. To obtain $\hat {\tilde \Gamma}_{1(2)}$ we multiply $\hat \Gamma_{1(2)}$ by the operator $\{1+[(\chi+\chi^{int}_{1(2)}) R_{1(2)}/\hbar]\bm {\sigma}({\bf \hat p}\times {\bf n})- [2m^*\gamma_{c_{1(2)}}^{int}/\hbar^3](\sigma_y p_y - \sigma_x p_x)\}^{-1}$  from the left and neglect the terms nonlinear over the SOI parameters.

If the system allows to perform spin diagonalization, the operators $\hat g_1$ and $\hat g_2$ transform into scalar quantities $\Delta d_1$ and $\Delta d_2$, respectively, having the dimensionality of length. In this case the operator $\hat \Gamma_{1(2)}$ just shifts the right (left) boundary to the new position $z_r=d/2+\Delta d_1$ ($z_l=-d/2+\Delta d_2$) which depends on the spin projection $\sigma=\pm 1$ and the corresponding interfacial parameters. Such spin diagonalization is possible in the systems with only one type of SOI (Rashba or Dresselhaus).

We begin with the case when only Rashba SOI is present. The resulting problem reads as

\begin{equation}
\label{H_onlyRashba}
\left(\frac{\hat p_z^2}{2m^*}+V(z)+a_{so}p\sigma \partial_z V\right)\psi_{\sigma}(z)=E_{\sigma}\psi_{\sigma}(z),
\end{equation}

\begin{equation}
\label{BC_onlyRashba}
\psi_{\sigma}(z)|_{z=d/2+\Delta d_1}=0, \qquad \psi_{\sigma}(z)|_{z=-d/2+\Delta d_2}=0,
\end{equation}

\begin{equation}
\label{Delta_d_onlyRashba}
\Delta d_{1(2)}=-R_{1(2)}n_z+\frac{\tilde{\chi}_{1(2)}R^2_{1(2)}}{\hbar}p\sigma,
\end{equation}
where $\tilde{\chi}_{1(2)}=\chi+\chi^{int}_{1(2)}$, $p$ is the absolute value of 2D momentum.

The further analysis is organized as follows. At first we consider the simple problem

\begin{equation}
\left(\frac{\hat p_z^2}{2m^*}+V(z)\right)\psi^{(0)}=E^{(0)}\psi^{(0)}
\end{equation}

\begin{equation}
\psi^{(0)}(z)|_{z=z_r}=0, \qquad \psi^{(0)}(z)|_{z=z_l}=0,
\end{equation}
which allows an exact numerical solution for an arbitrary potential profile $V(z)$. Next we assume $\Delta d_{1}$ and $\Delta d_{2}$ to be much smaller than $d$ and obtain the energy spectrum of the problem (\ref{H_onlyRashba})--(\ref{Delta_d_onlyRashba}). In the lowest order over the SOI parameters it reads as

\begin{multline}
E_{\sigma}=E^{(0)}+a_{so}\langle \psi^{(0)} |\partial_ z V |\psi^{(0)} \rangle p \sigma +\left.\frac{\partial E^{(0)}}{\partial z_r}\right|_{-\frac{d}{2},\frac{d}{2}} \Delta d_1 +\\+\left.\frac{\partial E^{(0)}}{\partial z_l}\right|_{-\frac{d}{2},\frac{d}{2}} \Delta d_2.
\end{multline}

Finally, we calculate the spin splitting $E_{SS}=(E_{+1}-E_{-1})$ and, comparing it with (\ref{NQW_SS}), obtain $\alpha_{SIA}$

\begin{equation}
\label{alpha_SIA_general}
\alpha_{SIA}=\alpha_{SIA}^{(0)} +\left.\frac{\tilde{\chi}_1R^2_1}{\hbar}\frac{\partial E^{(0)}}{\partial z_r}\right|_{-\frac{d}{2},\frac{d}{2}} +\left.\frac{\tilde{\chi}_2R^2_2}{\hbar}\frac{\partial E^{(0)}}{\partial z_l} \right|_{-\frac{d}{2},\frac{d}{2}}.
\end{equation}

Here the last two terms are the desired interfacial contributions. 

In the system with only Dresselhaus-type SOI we have 

\begin{equation}
\label{Delta_d_onlyDress}
\Delta d_{1(2)}=-R_{1(2)}n_z-\frac{2m^*\tilde{\gamma}_{c_{1(2)}}R_{1(2)}}{\hbar^3}n_z p\sigma,
\end{equation}
where $\tilde{\gamma}_{c_{1(2)}}=\gamma_c+\gamma_{c_{1(2)}}^{int}$. Performing the similar analysis as before we obtain interfacial contributions to $\alpha_{BIA}$

\begin{multline}
\label{alpha_BIA_general}
\alpha_{BIA}=\alpha_{BIA}^{(0)} -\left.\frac{2m^*(\gamma_c+\tilde{\gamma}_{c_{1}})R_1}{\hbar^3}\frac{\partial E_1^{(0)}}{\partial z_r}\right|_{-\frac{d}{2},\frac{d}{2}}-\\-\left.\frac{2m^*\tilde{\gamma}_{c_{1}}R_1}{\hbar^3}\frac{\partial E_2^{(0)}}{\partial z_r} \right|_{-\frac{d}{2},\frac{d}{2}}+ \left.\frac{2m^*(\gamma_c+\tilde{\gamma}_{c_{2}})R_2}{\hbar^3}\frac{\partial E_1^{(0)}}{\partial z_l}\right|_{-\frac{d}{2},\frac{d}{2}}+\\+\left.\frac{2m^*\tilde{\gamma}_{c_{2}}R_2}{\hbar^3}\frac{\partial E_2^{(0)}}{\partial z_l}\right|_{-\frac{d}{2},\frac{d}{2}},
\end{multline}
where $E_1^{(0)}=\langle \psi^{(0)}(z)| \hat p_z^2 / 2m^* |\psi^{(0)}(z)\rangle$, $E_2^{(0)}=\langle \psi^{(0)}(z)| V(z) |\psi^{(0)}(z)\rangle$.

One can expect that since all spin-orbit constants are small in general case when both Rashba and Dresselhaus type terms are allowed in the effective spin Hamiltonian, the corresponding spin splitting has the form (\ref{NQW_SS}), where in the lowest order over SOI parameters, $\alpha_{SIA}$ and $\alpha_{BIA}$ are still determined by Eqs. (\ref{alpha_SIA_general}) and (\ref{alpha_BIA_general}), respectively. This assumption will be verified below on a simple example.

It is important to note that the interfacial contributions to $\alpha_{SIA}$ and $\alpha_{BIA}$ can be calculated in a QW with arbitrary doping level and potential distribution since one is always able to find $E^{(0)}$ and $\psi^{(0)}$ numerically. However, in some cases transparent analytical results can be obtained. As an example we now consider the "narrow" QW in which the size quantization energy much exceeds the energy of the electron interaction with the smooth (in the atomic scale) potential $V(z)$. Treating the potential $V(z)$ as a perturbation, we obtain from Eqs. (\ref{alpha_SIA_general}) and (\ref{alpha_BIA_general}) for the ground subband 

\begin{multline}
\label{NQW_SIA}
\alpha_{SIA}=\alpha_{SIA}^{(0)}-\frac{2\tilde{E}_0}{\hbar} \frac{(\tilde{\chi}_1R_1^2-\tilde{\chi}_2R_2^2)}{d}+\\+\frac{eFd}{2E_0}\frac{E_0}{\hbar} \frac{(\tilde{\chi}_1R_1^2+\tilde{\chi}_2R_2^2)}{d},
\end{multline}

\begin{multline}
\label{NQW_BIA}
\alpha_{BIA}=\alpha_{BIA}^{(0)}+\frac{2k_0^2}{\hbar}\biggl[\frac{\gamma_c(R_1+R_2)}{d}+\\+\frac{\tilde{E}_0}{E_0}\frac{(\tilde{\gamma}_{c1}R_1+\tilde{\gamma}_{c2}R_2)}{d}+\frac{eFd}{4E_0}\frac{(\tilde{\gamma}_{c2}R_2-\tilde{\gamma}_{c1}R_1)}{d}\biggr],
\end{multline}

where $k_0=\pi/d$, $eF=\langle \psi_0(z)|V'(z)|\psi_0(z)\rangle$, $\psi_0(z)=\sqrt{2/d}\cos k_0z$,

\begin{equation}
E_0=\frac{\hbar^2k_0^2}{2m^*}, \qquad   \tilde{E}_0=E_0-\frac{1}{2}\langle \psi_0(z)|zV'(z)|\psi_0(z)\rangle, 
\end{equation}

It follows from Eqs. (\ref{NQW_SIA}) and (\ref{NQW_BIA}) that the ISOI not only renormalizes the values of $\alpha_{BIA}$ and $\alpha_ {SIA}$, but also affect the qualitative behaviour of the spin splitting. The spin splitting is anisotropic in the structures where both Dresselhaus- and Rashba-type SOIs are present. In the framework of the envelope functions approximation, $\alpha_ {SIA}$ is commonly assumed to be nonzero only in the structures with built-in or external electric field. However, it is seen from the Eq. (\ref{NQW_SIA}) that if the interfaces are non-equivalent, i. e. $\tilde{\chi}_1R_1^2 \neq \tilde{\chi}_2R_2^2$, the  $\alpha_ {SIA}$ is finite even in the QWs with zero average electric field. This effect gives rise to anisotropy of the spin splitting in such structures. Our theory naturally explains the results of Ref.[\onlinecite{Koralek_Nature_2009}] where significantly nonzero $\alpha_ {SIA}$ was observed in the nominally symmetric QW with equally doped sides and zero average electric field. At the same time, the interfacial contribution to $\alpha_{BIA}$ is nonzero even for structures with identical boundaries, i.e. $\gamma_{c_1}^{int}=\gamma_{c_2}^{int}$  and $R_1=R_2$. 

Now we check if the above assumption regarding the additivity of the Rashba and Dresselhaus contributions in the lowest order over SOI constants is fulfilled in the "narrow" QW. For this purpose, we calculate the spin splitting starting from the 3D problem (\ref{H})--(\ref{BC}) with both types of SOI and compare obtained $\alpha_{SIA}$ and $\alpha_{BIA}$ with ones satisfying (\ref{NQW_SIA}) and (\ref{NQW_BIA}). 

Aiming to analyse the effect of ISOI on the spin splitting, we take into account the interaction with the interfacial potential exactly. At the same time, bulk SOI $\hat H_{BIA}+\hat H_{SIA}$ and the smooth potential $V(z)$ which average value is assumed to be small in comparison with the size quantization energy are treated perturbatively. At first, we consider the following problem 
\begin{equation}
\label{basis1}
\frac{\hat p_z^2}{2m^*} \phi^{(0)}(z)=\epsilon^{(0)}\phi^{(0)}(z),
\end{equation}
\begin{equation}
\label{basis2}
\hat \Gamma_1 \phi^{(0)}(z)|_{z=d/2}=0,  \qquad   \hat \Gamma_2 \phi^{(0)}(z)|_{z=-d/2}=0.
\end{equation}

Introducing the values $\alpha_{1(2)}=2m^*\tilde{\gamma}_{c_{1(2)}} R_{1(2)} p/\hbar^3$, $\beta_{1(2)}=\tilde{\chi}_{1(2)}R^2_{1(2)}p / \hbar$, $\Delta=(\alpha_1+\alpha_2)e^{-i\phi}-i(\beta_1-\beta_2)e^{i\phi}$, $\tilde{\Delta}=(\alpha_1-\alpha_2)e^{-i\phi}+i(\beta_1-\beta_2)e^{i\phi}$ we obtain the eigenvalues and eigenfunctions of the problem (\ref{basis1})--(\ref{basis2}) in the lowest order over the scalar contributions of the interfaces and the ISOI parameters $\epsilon^{(0)}_{\pm}=E_0\left[1+2(R_1+R_2)/d \pm 2|\Delta|/d \right]$,

\begin{equation}
\label{wf}
\phi_{\pm}^{(0)}(z)=\left( \begin{array}{cc}
       C_1^{\pm} e^{-ik_{\pm}z}+C_3^{\pm} e^{ik_{\pm}z} \\
         C_2^{\pm} e^{-ik_{\pm}z}+C_4^{\pm} e^{ik_{\pm}z} \\
    \end{array} \right),
\end{equation}

where $k_{\pm}=k_0\left[1+(R_1+R_2)/d \pm  |\Delta|/d\right]$, 

\begin{equation}
C_2^{\pm}=\mp \frac{\Delta^*}{|\Delta|}C_1^{\pm} , \qquad C_3^{\pm}=\biggl[1+ik_0(R_1-R_2) \pm \frac{\tilde{\Delta}}{\Delta}|\Delta| \biggr]C_1^{\pm},
\end{equation}

\begin{equation}
C_4^{\pm}=\biggl[\mp \frac{\Delta^*}{|\Delta|}-ik_0\tilde{\Delta}^*\biggr] C_1^{\pm}.
\end{equation}

and $|C_1|=(1/4d)\left[1-(R_1+R_2)/d \mp |\Delta| / d\right]^{-2}$.

Next we find the spectrum of the problem (\ref{H})--(\ref{BC}) using the eigenfunctions  (\ref{wf}) as a basis $\phi(z)=A\phi^{(0)}_+ +B\phi^{(0)}_-$ and considering $\hat \delta H=\hat H_{BIA}+\hat H_{SIA}+V(z)$ as a perturbation 

\begin{multline}
\left( \begin{array}{cc}
      \epsilon^{(0)}_+ +\langle \phi^{(0)}_+ | \delta \hat H | \phi^{(0)}_+ \rangle  &\langle \phi^{(0)}_+ | \delta \hat H | \phi^{(0)}_- \rangle \\
       \langle \phi^{(0)}_- | \delta \hat H | \phi^{(0)}_+ \rangle  &  \epsilon^{(0)}_- + \langle \phi^{(0)}_- | \delta \hat H | \phi^{(0)}_- \rangle \\
    \end{array} \right)   \left( \begin{array}{cc}
      A \\
       B \\
    \end{array} \right)= \\=E \left( \begin{array}{cc}
      A \\
       B \\
    \end{array} \right).
\end{multline}

As it was expected, we obtain that the resulting spin splitting of the spectrum has the form (\ref{NQW_SS}) with $\alpha_{SIA}$ and $\alpha_{BIA}$ satisfying Eqs. (\ref{NQW_SIA}) and (\ref{NQW_BIA}), respectively. 

The values of the interfacial parameters can be extracted from comparison with the experiment, as it was done for the wide one-side doped quantum well GaAs/AlGaAs in Ref.[\onlinecite{ Devizorova_PhETF_2014}]. Let us, for example, evaluate the values $\tilde{\chi}_1R_1^2$ and $\tilde{\chi}_2R_2^2$ comparing Eq. (\ref{NQW_SIA}) with the experimental data from Ref.[\onlinecite{Koralek_Nature_2009}]. Due to the fact that all quantum wells were grown under the same conditions we assume the interfacial parameters to be equal for all structures. However, in each quantum well the left interface is not equivalent to the right one. We also suppose  $\tilde{E}_0 \approx E_0$. For the symmetrical sample with $F=0$ and $d=12$ nm the parameter $\alpha_{SIA}=(0.4 \times 10^{-3}) v_F$ ($v_F=4.11 \times 10^7$ sm/s is the Fermi velocity) is determined from the difference between $\tilde{\chi}_1R_1^2$ and $\tilde{\chi}_2R_2^2$ . Thus, we can estimate this difference as $(\tilde{\chi}_1R_1^2 - \tilde{\chi}_2R_2^2)=1.4$ \AA$^2$. Next we consider the sample with the same thickness and asymmetrical doping for which \cite{Koralek_Nature_2009} $\alpha_{SIA}=(1.3 \times 10^{-3}) v_F$. Performing the self-consistent solution of the Shrodinger and Poisson equations, we obtain $F=2.085 \times 10^5$ V/sm. Thus, we evaluate  $(\tilde{\chi}_1R_1^2 + \tilde{\chi}_2R_2^2)=7.7$ \AA$^2$. Finally, we find $\tilde{\chi}_1R_1^2=4.6$ \AA$^2$, $\tilde{\chi}_2R_2^2=3.1$ \AA$^2$. The experimental data presented in Ref.[\onlinecite{Koralek_Nature_2009}] is not enough to calculate $\tilde{\chi}_{1}$, $\tilde{\chi}_{2}$, $R_1$ and $R_2$ separately. However, some estimates can be obtained. The typical value of $R$ is in the order of $\sim 20$ \AA \cite{Devizorova_PhETF_2014}. Thus, we evaluate  $\tilde{\chi}_{1} \sim 0.012$ and $\tilde{\chi}_{2} \sim 0.008$. The corresponding values $\chi^{int}_1 \sim -0.07$ and $\chi^{int}_2 \sim -0.74$ are the same order as bulk value $\chi=0.082$.

In conclusion we developed the theory of ISOI in the narrow QWs. We have obtained the renormalization of the Dresselhaus and Rashba parameters arising from the SOI at two heterointerfaces. The considerable spread in the experimentally determined values of spin-orbit constants can originate from the dependence of  $\alpha_{BIA}$ and $\alpha_{SIA}$ on the interfacial parameters and, thus, on the growth conditions. We also have demonstrated that the microscopic dissimilarity of the interfaces leads to the finite Rashba parameters even in the QWs with zero average electric field. This result explains the experimental data of Ref.[\onlinecite{Koralek_Nature_2009}] where nonzero $\alpha_{SIA}$ was obtained in the symmetric structure. 

This work is supported by the state assignments AAAA-A16-116041410063-1.  Zh.A.D. is supported by the Russian Foundation for Basic Research (project 16-32-00708) and the Dynasty Foundation.

\end{document}